\documentclass{esannV2}
\usepackage[dvips]{graphicx}
\usepackage[latin1]{inputenc}
\usepackage{amssymb,amsmath,array}

\begin{document}
\title{Decoding Finger Flexion using Amplitude Modulation from Band-specific ECoG}

\author{Nanying Liang$^1$ and Laurent Bougrain$^2$
%
%
\vspace{.3cm}\\
%
1- CORTEX INRIA project-team, INRIA-LORIA \\
615 rue du Jardin Botanique, Villers-l\`{e}s-Nancy, 54600 France
%
\vspace{.1cm}\\
2- CORTEX INRIA project-team, Nancy University-LORIA \\
campus scientifique, BP 239, Vandoeuvre-l\`{e}s-Nancy, 54506 France \\
}

\maketitle

\begin{abstract}
EEG-BCIs have been well studied in the past decades and implemented into several famous applications, like P300 speller and wheelchair controller. However, these interfaces are indirect due to low spatial resolution of EEG. Recently, direct ECoG-BCIs attract intensive attention because ECoG provides a higher spatial resolution and signal quality. This makes possible localization of the source of neural signals with respect to certain brain functions. In this article, we present a realization of ECoG-BCIs for finger flexion prediction provided by BCI competition IV. Methods for finger flexion prediction including feature extraction and selection are provided in this article. Results show that the predicted finger movement is highly correlated with the true movement when we use band-specific amplitude modulation.
\end{abstract}

\section{Introduction}

A brain-computer interface (BCI) is a new communication and control option for those with severe motor disabilities \cite{Wolpaw2002}. A BCI system translates brain signals into commands for a computer or other devices. Electroencephalography (EEG) based BCIs are the most studied non-invasive interfaces mainly due to the fine temporal resolution, ease of use, portability and low set-up cost of EEG recordings \cite{Hughes1994}. Unfortunately, non-invasive implants produce noisy signals because of the damping effect introduced by the skull. Further, its spatial resolution is low due to the size of the electrode. Thus, it is difficult to use EEG signals for directly decoding the task events.

Electrocorticography (ECoG) has recently emerged as a promising recording technique for use in brain-computer interfaces (BCI) and has been extensively investigated \cite{Schalk2007,Sanchez2008,Chin2007,Pistohl2008}. ECoG electrodes arrays were initially implanted (under the skull but over the surface of the cortex) for severe epileptic patients for presurgical planning, in order to identify the sources generating epileptic seizures \cite{Asano2005}. The spatial resolution of ECoG signals is higher than EEG with a better signal-to-noise ratio (SNR). Typically, the diameter of one ECoG electrode is of 4 mm with 1 cm inter-electrode distance. Therefore ECoG can provide a spatial resolution of approximately 1 cm \cite{Asano2005}. Spatial resolution plays an important role in BCI \cite{Sanchez2008}. The fine spatial resolution of ECoG provides an opportunity for decoding brain functions directly and therefore possibly leads to the implementation of direct neural interfaces, which is difficult to be accomplished through EEG-based BCIs.

In order to study the application of ECoG in BCIs, several research groups have recorded ECoG signals when the participants performed certain kind of tasks related to the brain functional areas that the implanted electrode arrays had covered. The testing tasks include center-out reaching or pointing task \cite{Sanchez2008}, finger flexion \cite{Schalk2007} and cursor trajectory \cite{Pistohl2008}. Some of these groups are interested in the use of frequency features as the inputs of decoder \cite{Pistohl2008}, and others use amplitude modulation \cite{Sanchez2008}, or both \cite{Schalk2007}. For features extracted from frequency domain, it was found that different frequency bands have different correlation coefficients to the event tasks \cite{Schalk2007,Chin2007}. Further, the implanted ECoG electrode array is usually a grid with a size from 6*6 to 8*8. It covers a cortical zone involving several different functions of the brain. Thus, frequency band and electrode selection are necessary for identifying good descriptors.

In this paper, we first present the material used, the ECoG data set adopted from BCI competition IV in section \ref{Material}. The methods for feature extraction and selection are presented in \ref{Methods-1}. The construction of a linear regression model for decoding is done in \ref{Methods-2}. Simulation results on this competition data set are reported in section \ref{Results}. We draw conclusions with discussions on further work in section \ref{Conclusion and Further Work}.

\section{Material}\label{Material}
\begin{figure}[h!]
\centering
\includegraphics[scale=0.5]{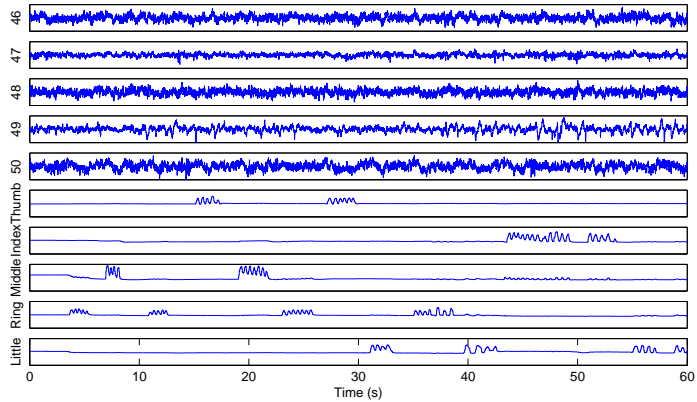}
\caption{Example ECoG signals from channel 46 to 50 (the first 60 seconds in the training data set) from subject 1. The last 5 rows show the corresponding finger movement time course.}
\label{Fig:data_visulization}
\end{figure}

Recently several paradigms of ECoG-BCIs have been realized focusing on motor tasks. Here, we just adopt the ECoG-BCI data set provided by BCI competition IV \cite{Schalk2007,Miller2008}. Three subjects participated in the recording were epileptic patients. Each subject was implanted a subdural electrode array with  the arrangement of 8*6 or 8*8 electrodes. The ECoG signals were recorded when the subjects performed a finger movement task. The subjects were instructed to move one certain finger by the corresponding word displayed on a computer screen. The execution of finger movement lasted 2 seconds and it was followed by a 2-second resting period. There were 30 movement stimulus for each finger resulting in 600-second recording for each subject. The ECoG signals were recorded through the general-purpose BCI system BCI2000 \cite{Schalk2004}, bandpass filtered between 0.15 to 200 Hz and sampled at 1000 Hz. The finger movements were recorded using a dataglove sampled at 25 Hz. Figure \ref{Fig:data_visulization} shows an example of the visualization of the ECoG signals and the finger movement time course from subject 1. Due to space limitation, only a subset of electrodes is displayed.

\section{Decoding Workflow}\label{Methods}
\subsection{Feature Extraction and Selection}\label{Methods-1}
The evidence of sensorimotor ECoG dynamics has been reported in several specific frequency bands including sub-bands (1-60 Hz), gamma band (60-100 Hz), fast gamma band (100-300 Hz) and ensemble depolarization (300-6k Hz) \cite{Sanchez2008}. Being inspired by the firing rate coding scheme used in spike train decoding, Sanchez proposed the band-specific amplitude modulation (AM) as the descriptor for ECoG signal decoding, which is defined as the sum of the power of the voltage of the band-specific ECoG signals in a time bin $\Delta t$:

\begin{equation}\label{Equ:AM}
    x(t_{n}) = \sum^{\Delta t}_{t=0}v^{2}(t_{n}+t)
\end{equation}

where $t_{n+1} = t_{n} + \Delta t$. We simply let $\Delta t = 40 ms$ so the resulting band-specific AM feature inputs have the same sampling rate (i.e. 25Hz) as that of the dataglove position measurements. It seems that the sensitivity profile of ECoG for each frequency band is specific to a given task \cite{Sanchez2008}. For our study, we also found that each finger movement was correlated to two or three specific electrodes. Therefore, feature selection including frequency band and electrode site selection is necessary for removing the unrelated input descriptors. For each finger and each subject, we use a stepwise feature selection procedure based on the method of train and validation (i.e. 3/5 of available data are used for training and 2/5 for validation).
In the future, we will refer to this method as FD-ECoG AM for frequency decomposition with feature extraction scheme.

Next, we want to compare it with other methods for estimating AM features: raw-ECoG AM feature and PCA-ECoG AM feature. For computing the raw-ECoG AM feature, we repeat the above procedure but exclude the frequency decomposition operation. For PCA-ECoG AM feature, we first apply the principal component analysis (PCA) method to the whole ECoG data set, and then for each principal component, we calculate the corresponding AM feature. After obtaining the descriptor features, we are ready to present a decoder for predicting finger movement in the following section.

\subsection{Linear Regressor}\label{Methods-2}
We have no a priori information about the relationship between the descriptors and the target signals or the interaction between descriptors. Thus, we simply applied a linear model as a decoder for its robustness property. We also noticed that other advanced methods have been used for ECoG signals decoding, for example, the Kalman filter \cite{Pistohl2008} and nearest neighbour classifier (NNC) \cite{Chin2007}. But these methods are not suited for the case being studied here because the first method needs a finger model which we do not have and the second one is suitable for classification. Linear models have been applied to ECoG signal analysis by other research groups with successful outcome \cite{Schalk2007,Sanchez2008}. Here, we adopt the linear model with the form as follows:
\begin{equation}\label{Equ:Wiener}
d(t) = W^{T}x(t)
\end{equation}

where $d$ is the finger position as measured by a dataglove. $x(t)$ is the tap-delay AM feature vector. The coefficients $W$ of the model are trained by the Wiener solution in Equation (\ref{Equ:Wiener}):

\begin{equation}\label{Equ:Wiener}
W = E(x^{T}x)^{-1}E(x^{T}d)
\end{equation}

where $E$ is the expected mean. The number of tap-delay is optimized with the value of 25 tap-delays for our case (i.e., using the present AM input and the previous 1-second AM inputs for predicting the present finger flexion). Further, in order to improve the stability for estimating the coefficients of the Wiener model, we replace the inverse operation in Equation (\ref{Equ:Wiener}) with the pseudo-inverse.

\section{Results}\label{Results}
First, we present the simulation results of the linear decoder combined with different feature extraction methods as summarized in Table \ref{Tab:PC-preprocessing}\footnote{The performance is evaluated only with the validation data set because the finger position measurements because the testing data set are not available yet by the BCI competition IV.}.

\begin{table}[h!]
  \centering
  \begin{tabular}{|c|c|c|c|c|c|c||c|}
    \hline
    Subj. & AM feature & Thumb & Index & Middle	& Ring & Little & Av. \\
    \hline
     & Raw ECoG AM & 0.26 & 0.26 & 0.25	& 0.40 & 0.32 & 0.30 \\
    1& PCA-ECoG AM & 0.22 & 0.29 & 0.22 & 0.33 & 0.23 & 0.26 \\
     & FD-ECoG AM  & 0.56 & 0.76 & 0.37 & 0.62 & 0.48 & 0.56 \\
    \hline
     & Raw ECoG AM & 0.46 & 0.31 & 0.38 & 0.41 & 0.25 & 0.36 \\
    2& PCA-ECoG AM & 0.30 & 0.25 & 0.28 & 0.27 & 0.19 & 0.26 \\
     & FD-ECoG AM  & 0.64 & 0.40 & 0.50 & 0.57 & 0.45 & 0.51 \\
    \hline
     & Raw ECoG AM & 0.63 & 0.46 & 0.54 & 0.56 & 0.39 & 0.52 \\
    3& PCA-ECoG AM & 0.31 & 0.34 & 0.44 & 0.42 & 0.35 & 0.37 \\
     & FD-ECoG AM  & 0.73 & 0.68 & 0.78 & 0.68 & 0.64 & 0.70 \\
    \hline
  \end{tabular}
  \caption{Performance of the linear decoder combined with different feature extraction methods in terms of correlation coefficient between the predicted and true finger movement for each finger and subject. The last column presents the averaged result for each method and subject.}\label{Tab:PC-preprocessing}
\end{table}

\begin{figure}[h!]
\begin{center}
\includegraphics[scale=0.65,viewport=685 0 0 305]{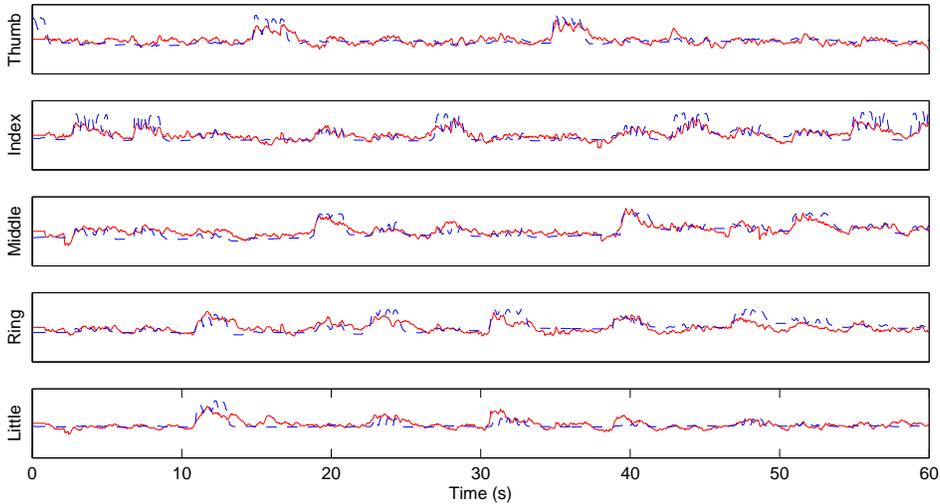}
\caption{Predicted (solid red) and real (dash blue) finger movement time courses for the first 60 seconds in validation data set from subject 3.}
\label{Fig:performance_visulization}
\end{center}
\end{figure}

Table \ref{Tab:PC-preprocessing} shows that the method based on AM features generated from band-specific ECoG signals outperforms the others. It can be inferred that the sensitivity profile of ECoG is both band-specific and channel-specific to a task and subject. Therefore feature selection is important for the ECoG signal decoding. For example, in the sense of correlation, channel 1 ranks highest for flexion decoding for thumb and index finger, and channel 39 ranks highest for middle, ring and little finger for subject 1. We provide an example for the predicted finger movement for subject 3 based on the method of FD-ECoG AM as shown in solid red curve in Figure \ref{Fig:performance_visulization}. For comparison purpose, the corresponding true finger movement time course is plotted in the dash blue curve.

\section{Conclusion and Further Work}\label{Conclusion and Further Work}
In this paper, we have studied a linear decoding scheme for predicting finger movement from ECoG signals. The approach presented in this article won the BCI competition IV addressed to this problem. The experimental results show that, with the fine spatial resolution of ECoG signals, it is promising to implement a direct neural interface in applications, for example, dedicated robotic arm/hand control and simple gesture language communication such as cued language. Moreover, when comparing the results obtained by the method taking into account the frequency-specific features and those not, it can be inferred that the sensitivity profile of ECoG signals is band-specific and channel-specific to a task and subject.

In the stepwise feature selection procedure, we did not consider the correlation between band-specific ECoG signals. It is suggested that incorporating the feature correlation into feature selection, for example, using correlation feature selection (CFS) method \cite{Hall1999}, may produce an optimal compact feature set.

We have no a priori information about the relationship between the descriptors and the target signals or the interaction between descriptors and thus make a simple assumption of linear regression between the descriptors and the target signals. We try to introduce some user-defined interaction terms into the regression model. It may improve the decoding accuracy using less number of regressors. Thus, non-parameter models incorporating the interaction terms into the decoding model, for example, using multivariate adaptive regression splines (MARS) \cite{Friedman1991} will be investigated in the future.


\begin{footnotesize}





\end{footnotesize}


\end{document}